\documentstyle[12pt,epsf]{article}
\begin{document}

\thispagestyle{empty}
\begin{center}

{\large {\bf Parametric X-radiation for Bragg geomery.}}
\medskip

{Institute of Nuclear Problems, Belarussian State University, Minsk}

\bigskip

\end{center}

\section*{Introduction}

As it is well known the refraction of the electromagnetic field associated with
an electron passing through a matter with a uniform velocity originates
the Vavilov-Cherenkov and transition radiation. The Cherenkov emission of
photons by a charged particle occurs whenever the index of refraction $%
n(\omega )>1$ ($\omega $ is the photon frequency). For the X-ray frequencies
being higher than atom's ones the refractive index has the general form
\begin{equation}
n(\omega )=1-\frac{\omega _{L}^{2}}{2\omega ^{2}},
\end{equation}
where $\omega _{L}$ is the Lengmuir frequency.

But as it was shown for the first time in \cite{1} 
in the X-ray region the crystal
structure essentially changes the refraction index for virtual photons
emitted under diffraction. This leads to the change of generation conditions
for bremsstrahlung and Cherenkov radiation. As a result, when a
charged particle is moving at a constant velocity it can emit
quasi-Cherenkov X-ray radiation \cite{1,2}. Moreover, the particle can emit
photons not only at the angle $\vartheta \sim \gamma ^{-2}$ 
($\gamma $ is Lorentz factor of the particle) but at the
much larger angle as well and
even in the backward direction. 
This proved to be the important circumstance
that allowed to observe Parametric X-ray Radiation (PXR) for the first time
\cite{3,4}. The observed radiation peaks were shown \cite{3,5,6} to
be really a new type of radiation and not diffraction of bremsstrahlung
radiation.

In accordance with [2-4] the diffraction of virtual photons in crystal 
can be described by a set of
refraction indices $n_{\mu }(\omega ,\vec{k})$, some of which
may appear to be greater than unit. 
Here $n(\omega ,\vec{k})$ is the refraction index of crystal for X-ray
with photon wave vector $\vec{k}$ . Particularly, in the case of two-wave
diffraction the refraction index $n_1 (\omega ,\vec{k})>1$ and
$n_2 (\omega ,\vec{k})<1$, and, respectively, 
two waves propagate in crystal -- the fast one ($n_{2}<1$) and the slow
one ($n_{1}>1$). For the slow wave the Cherenkov condition can be fulfilled
\begin{equation}
1-\beta n(\omega ,\vec{k})\cos \vartheta =0,
\end{equation}
where \ $\vartheta $\ is the angle between photon wave
vector $\vec{k}$ and the particle velocity $\vec{v}$ (polar angle of
radiation), $\beta =v/c,$ $c$ is the velocity of light. As a result, this
wave can be emitted all along the particle pass in crystal.
Condition (2) is not valid for the fast wave, and this wave can be emitted
only at the  vacuum-crystal boundary. When the diffraction conditions are not 
fullfilled  the well-known ordinary transition X-ray radiation (TXR) is
formed.

PXR is highly directed, quasimonochromatic, tunable, polarized radiation with
high spectral intensity. These properties as well as the 
possibility of radiation at
the angles being much greater than typical radiation angle of a relativistic
particle $\gamma ^{-1}$ make PXR potentially useful for wide variety of
applications such as materials analysis, microscopy, nanofabrication and
digital subtraction imaging. Predictions for the spectral and angular
features of PXR have been corroborated by \cite{5,6,13,14}.

In the above referenced papers the detailed analysis for a
diffraction maximum at the angle $\vartheta >>\gamma ^{-1}$ 
has been performed.
So in some
contributions it has been concluded that PXR is emitted at a large angle
only \cite{12,14} and it can be described
in the
kinematic approximation by Ter-Mikaelian formula \cite{15} for
propagation of radiation in a periodic medium. According to \cite{14} PXR is
the
well-known resonance radiation \cite{15}. According to \cite{14} there
is a new "PXR
of type B" for low energy electrons, which is not related to the Cherenkov
effect. 

However, it should be noted that according to Baryshevsky and Feranchuk
\cite{9,10}, each photon with the frequency $\omega $, emitted at a large angle,
corresponds to the photon with the same frequency $\omega $,
emitted at a small
angle $\vartheta \leq \gamma^{-1}$ near forward direction (particle's
velocity direction). The frequency of this photon does not depend on
particle frequency in contrast with resonance radiation photons, frequency of
which is proportional to $\gamma ^{2}$. Moreover, the PXR at small angles
cannot be described in the
kinematic approximation by Ter-Mikaelian formula in principle. 
The kinematic theory cannot describe
backward Bragg diffraction either.

Up to now PXR at small angle near the relativistic
particle velocity direction has not been observed experimentally.
PXR
observation in the forward diffraction maximum would uniquely prove out 
the quasi-Cherenkov mechanism of PXR.

In \cite{16} the expressions for angular distribution of PXR in forward
diffraction maximum  neglecting absorption were derived. For Laue
geometry the total number of photons in forward maximum was obtained.

The assumption of low absorption does not allow to take into account the
possible total reflection of quanta in crystal at Bragg geometry.
Because even at negligibly small ordinary absorption, the rapidly damped 
inside the crystal wave  
arises (so called the internal extinction phenomenon
\cite{17}). In the present paper PXR for 
both  the forward and backward diffraction in Bragg geometry was studied.
It was shown that contributions of both the slow and the fast waves
in radiation
intensity are comparable and only their joint account allows to describe the
experimental results. The comparison of the theory and
experiment was carried out. The calculations  for
Si and LiH crystals was performed.

\section{General expressions for PXR emission rate.}

Let a particle moving with a uniform velocity be incident on a crystal plate
with the thickness $L$, $L<<L_{br}$   (where
\ $L_{br}=(\frac{\omega q}{c})^{-1/2}$\ is the coherent length of bremsstrahlung radiation, $q=\overline{\theta }%
_{s}^{2}/4$\ and $\overline{\theta }_{s}^{2}$ is a root-mean-square angle of
multiple scattering of charged particles per unit length). This
requirement allows us to neglect the multiple scattering of particles by
atoms of crystal. The theoretical method for case of intense 
multiple scattering can be found
in \cite{7}.

The expression for the differential number of photons with the polarization
vector $\vec{e}_{s}$ emitted in the $\vec{k}$ direction was obtained in,
for example, \cite{16}:
\begin{equation}
\frac{d^{2}N_{s}}{d\omega d\Omega }=\frac{e^{2}Q^{2}\omega }{4\pi ^{2}\hbar
c^{3}}\left| \int\limits_{-\infty }^{+\infty }\vec{v}\vec{E}_{\vec{k}%
}^{(-)s}(\vec{r}(t),\omega )\exp (-i\omega t)dt\right| ^{2},
\end{equation}
where $eQ$ is the particle charge, $\vec{E}_{\vec{k}}^{(-)s}$\ is the
solution of homogeneous Maxwell's equation. In order to determine the number
of quanta emitted by a particle passing through a crystal plate one should
find the explicit expressions for the solutions $\vec{E}_{\vec{k}%
}^{(-)s}$ . The field $\vec{E}_{\vec{k}}^{(-)s}$ can be found from the
relation $\vec{E}_{-\vec{k}}^{(-)s}=(\vec{E}_{\vec{k}}^{(+)s})^{\ast }$\
using the solution $\vec{E}_{\vec{k}}^{(+)s}$, that describes the ordinary
problem of photon scattering by a crystal.

The following set of equations  for   wave amplitudes
in the case of two-beam diffraction can be obtained:
\begin{eqnarray}
\left( \frac{k^{2}}{\omega ^{2}}-1-\chi _{0}^{\ast }\right) \vec{E}_{\vec{k}%
}^{(-)s}-C_{s}\chi _{-\tau }^{\ast }\vec{E}_{\vec{k}_{\tau }}^{(-)s} &=&0
\nonumber \\
\left( \frac{k^{2}}{\omega ^{2}}-1-\chi _{0}^{\ast }\right) \vec{E}_{\vec{k}%
}^{(-)s}-C_{s}\chi _{-\tau }^{\ast }\vec{E}_{\vec{k}_{\tau }}^{(-)s} &=&0.
\end{eqnarray}
Here $\vec{k}_{\tau }=\vec{k}+\vec{\tau}$ , $\vec{\tau}$ is the reciprocal
lattice vector ($|\tau |=2\pi /d,$ $\ d$ is the interplanar distance), $\chi
_{0},\chi _{\tau }^{s},\chi _{-\tau }^{s}$ are the Fourier components of the
crystal susceptibility. It is well known that a crystal is described by a
periodic susceptibility (see, for example \cite{18}):\bigskip
\begin{equation}
\chi (\vec{r})=\sum_{\tau }\chi _{\tau }\exp (i\vec{\tau}\vec{r}).
\end{equation}
$C_{s}=\vec{e}_{s}\vec{e}_{\tau s}$ , $\vec{e}_{s}$ $(\vec{e}_{\tau s})$ are
the unit polarization vectors of the incident and diffracted waves,
respectively.

The condition for the linear system (4) to be solvable leads to a dispersion
equation that determines the possible wave vectors $\vec{k}$ in a crystal.
It is convenient to present these wave vectors in the form$\ \ \ \vec{k}%
_{\mu s}=\vec{k}+\kappa _{\mu s}\vec{N},$ \ \ \ \ $\kappa _{\mu s}=\frac{%
\omega }{c\gamma _{0}}\varepsilon _{\mu s},$ where $\mu =1,2$; $\vec{N}$ is
the unit vector normal to the entrance crystalline plate surface and directed inward a crystal,
\begin{equation}
\varepsilon _{\mu s}=\frac{1}{4}\left\{ -\alpha _{B}\beta _{1}+\chi
_{0}(\beta _{1}+1)\pm \sqrt{\left[ -\alpha _{B}\beta _{1}+\chi _{0}(\beta
_{1}-1)\right] ^{2}+4\beta _{1}\chi _{\tau }^{s}\chi _{-\tau }^{s}}\right\},
\end{equation}
\begin{equation}
\alpha _{B}=\frac{2\vec{k}\vec{\tau}+\tau ^{2}}{k^{2}},
\end{equation}
$\alpha _{B}$ \ is the off-Bragg parameter ($\alpha _{B}=0$, if the exact
Bragg condition of diffraction is fulfilled), \ $\beta _{1}=\gamma
_{0}/\gamma _{1}$, $\gamma _{0}=\vec{n}_{\gamma }\vec{N}$, $\gamma _{1}=\vec{%
n}_{\gamma \tau }\vec{N}$, \ \ $\vec{n}_{\gamma }=\frac{\vec{k}}{k}$, $\vec{n%
}_{\gamma \tau }=\frac{\vec{k}+\vec{\tau}}{|\vec{k}+\vec{\tau}|}$.

The general solution of equation (4) inside a crystal is:
\begin{equation}
\vec{E}_{\vec{k}}^{(-)s}(\vec{r})=\sum_{\mu =1}^{2}\left[ \vec{e}^{s}A_{\mu
}\exp (i\vec{k}_{\mu s}\vec{r})+\vec{e}_{\tau }^{s}A_{\tau \mu }\exp (i\vec
{k}_{\mu s\tau }\vec{r})\right] .
\end{equation}

Matching these solutions with the solutions of Maxwell's equation for the
vacuum area we can find the explicit form of $\vec{E}_{\vec{k}}^{(-)s}(\vec{r%
})$ throughout the space. It is possible to discriminate several types of
diffraction geometries, namely, Laue and Bragg schemes are the most
well-known.

Let us consider the Bragg case. 
An electromagnetic wave emitted in the forward direction as well as 
the diffracted one emitted by a charged particle and leaving the
crystal through the surface of the particle entrance can be observed. Fig.1
demonstrates the schemes for forward (a) diffraction maximum and diffraction
maximum at large angle equal to $2\theta _{B}$ (here $\theta _{B}$ -- Bragg
angle, $\sin \theta _{B}=\frac{|\vec{v}\vec{\tau}|}{\tau }$) with respect to
particle's velocity direction (b) in Bragg geometry.

\begin{figure}[tbp]
\epsfxsize = 10 cm
\centerline{\epsfbox{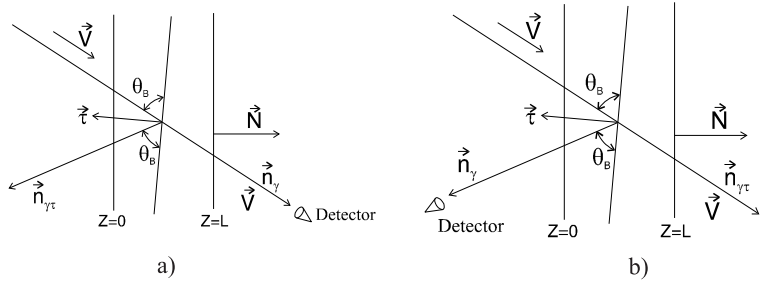}}
\caption{\it Schemes of diffraction geometries: a) Forward maximum, b)
Diffracted maximum.}
\end{figure}

Matching the solutions of Maxwell's equations on the crystal surface by use
(4), (6) and (8),
one can obtain the following expression:

$$
\vec{E}_{\vec{k}}^{(-)s}(\vec{r})=\vec{e}^{s}\left[ -\sum_{\mu
=1}^{2}\gamma _{\mu s}^{0\ast }e^{-i\frac{\omega }{c\gamma _{0}}
\varepsilon _{\mu s}^{\ast }L}\right] e^{i\vec{k}\vec{r}}\Theta(-z)+
$$

$$
\left\{ \vec{e}^{s}\left[ -\sum_{\mu =1}^{2}\gamma _{\mu s}^{0\ast }e^
{-i\frac{\omega }{c\gamma _{0}}\varepsilon _{\mu s}^{\ast }(L-z)}\right] e^
{i\vec{k}\vec{r}}+\vec{e}_{\tau }^{s} \beta _{1} \left[ \sum_{\mu=1}^{2}
\gamma _{\mu s}^{\tau \ast }e^{-i\frac{\omega }{c\gamma _{0}}
\varepsilon _{\mu s}^{\ast }(L-z)} \right] e^{i\vec{k}_{\tau }\vec{r}}
\right \}\Theta(L-z)\Theta (z)+
$$

\begin{equation}
\left\{ \vec{e}^{s} e^ {i\vec{k}\vec{r}}+\vec{e}_{\tau }^{s}\beta _{1}
\left[ \sum_{\mu =1}^{2}\gamma _{\mu s}^{\tau \ast }e^{-i\frac{\omega }{
c\gamma _{0}}\varepsilon _{\mu s}^{\ast }L}\right] e^ {i\vec{k}_{\tau }\vec{
r}}\right\}\Theta (z-L),
\end{equation}
where
\begin{eqnarray}
\gamma _{1(2)s}^{0} &=&\frac{2\varepsilon _{2(1)s}-\chi _{0}}{(2\varepsilon
_{2(1)s}-\chi _{0})-(2\varepsilon _{1(2)s}-\chi _{0})\exp (-i\frac{\omega }{c\gamma _{0}}
(\varepsilon _{2(1)s}-\varepsilon _{1(2)s})L)} \nonumber \\
\gamma _{1(2)s}^{\tau } &=&\frac{-\beta _{1}C_{s}\chi _{\tau }}{%
(2\varepsilon _{2(1)s}-\chi _{0})-(2\varepsilon _{1(2)s}-\chi _{0})\exp (-i%
\frac{\omega }{c\gamma _{0}}(\varepsilon _{2(1)s}-\varepsilon _{1(2)s})L)}.
\end{eqnarray}

According to (4) in case of two-beam diffraction
there are two different refraction indices for photons propagating
inside a crystal and one of them being able to become greater than unit. 
That is, the superposition
of slow and fast waves propagates in a crystal. 
The important thing is that the refraction indices for both waves
strongly depend on frequencies and propagation directions of photons.

Substituting (9) in (3) one can obtain the differential number of PXR quanta
with polarization vector
$\vec{e}_s$ diffracted forwards for the Bragg case:
\begin{equation}
\frac{d^2 N_s}{d \omega d\Omega }=\frac {e^2 Q^2 \omega }{4\pi^2
\hbar c^3}(\vec{e}_s \vec{v})^2
\left| \sum_{\mu =1,2}\gamma _{\mu s}^{0} e^{i\frac{\omega }{c\gamma _{0}}
\varepsilon _{\mu s}L}\left[ \frac{1}{\omega -\vec{k}\vec{v}}-\frac{1}
{\omega -\vec{k}_{\mu s}\vec{v}}\right]
\left[ e^{\frac{i(\omega -\vec{k}_{\mu s}\vec{v})L}
{c\gamma _{0}}}-1\right] \right| ^{2},
\end{equation}
here $\vec{e}_{1}||\left[ \vec{k}\vec{\tau}\right] $ , $\vec{e}_{2}||\left[
\vec{k}\vec{e}_{1}\right] $ , $\vec{k}_{\mu s}=\vec{k}+\frac{\omega }{%
c\gamma _{0}}\varepsilon _{\mu s}\vec{N}$.

For PXR quanta with the polarization vector $\vec{e}_{\tau s}$ diffracted at
a large angle relative to the charged particle velocity the above
formula can be rewritten as follows:
\begin{equation}
\frac{d^{2}N_{s}}{d\omega d\Omega }=\frac{e^{2}Q^{2}\omega }{4\pi
^{2}\hbar c^{3}}(\vec{e}_{\tau s}\vec{v})^{2}
\left| \sum_{\mu =1,2}\gamma _{\mu s}^{\tau }\left[ \frac{1}{\omega -\vec{k}_{\tau }
\vec{v}}-\frac{1}{\omega -\vec{k}_{\mu \tau s}\vec{v}}\right] \left[
e^{\frac{i(\omega -\vec{k}_{\mu \tau s}\vec{v})L}{c\gamma _{0}}}-1\right]
\right| ^{2}.
\end{equation}

\section{Transition radiation and PXR}

Let us compare the obtained formulae with the well-known ones for
x-ray radiation transition  in amorphous medium. The TXR formula is known
as follows:
\begin{equation}
\frac{d^{2}N_{s}}{d\omega d\Omega } =\frac{e^{2}Q^{2}\omega }{4\pi
^{2}\hbar c^{3}}(\vec{e}_{s}\vec{v})^{2}
\left| e^{i\frac{\chi _{0}\omega }{2c\gamma _{0}}L}\left[ \frac{1}{
\omega -\vec{k}\vec{v}}-\frac{1}{\omega -\vec{k}_{a}\vec{v}}\right] \left[
e^{\frac{i(\omega -\vec{k}_{a}\vec{v})L}{c\gamma _{0}}}-1\right] \right|
^{2},
\end{equation}
where $\vec{k}_{a}=\vec{k}+\frac{\omega \chi _{0}}{2c\gamma _{0}}\vec{N}$ \
is the photon wave vector in amorphous medium.
In amorphous medium or in a crystal away from the diffraction
conditions the only single wave	propagates.
But as a result of diffraction
a coherent
superposition of several waves propagates in a crystal.
One can see that (13) is very similar to the expression describing PXR.
When the frequencies and angles in (11) do not satisfy the
diffraction conditions, the formula for forward PXR turns to the TXR
one (13). At the same time the spectral-angular intensity
(12) appears to be equal to zero, that is quite obvious
since no diffraction at a large angle arises.

\section{PXR at small angles with respect to particle's velocity in Bragg
geometry.}

The expression for PXR angular distribution in forward direction can be
derived from (11) using some assumptions. First of all it should be noted that  $\chi _{0}<0$ and from (6) one can obtain that only
one root of $\varepsilon _{\mu s}$ \ gives the refractive index $n>1$. As a
result, the difference ($\omega -\vec{k}_{\mu s}\vec{v}$) for this $\mu $\
can be equal to zero and the term in (11) containing this difference 
will grow proportionally to $L$. On the other hand, the term in
(11) containing the same difference ($\omega -\vec{k}_{\mu s}\vec{v}$)
for the other root will not be proportional to $L$,
because this difference can never turn to zero. At the first glance it means
that the term containing the first difference as well as for the Laue case
gives the principle contribution to the radiation intensity
when the particle waylength in crystal $L_{0}$\ grows.
But the distinctive feature of
Bragg case is the possibility of total reflection of quanta that is
stipulated by existence of the nonhomogeneous wave in a crystal. So, as a
result, the both roots of $\varepsilon
_{\mu s}$ should be considered. Therefore, to obtain the PXR angular
distribution it is necessary to perform the numerical integration of (11)
over frequencies in the vicinity of $\omega _{B}$.

Assuming that \ $L_{0}>>l$\ ($l=\lambda \gamma ^{2}$ is the vacuum coherent
length, $\lambda $ is wave length of the photon) and neglecting the imaginary
parts of crystal susceptibilities and possible total reflection we can
obtain the approximate expression for PXR angular distribution in forward
direction:
\begin{eqnarray}
\frac{dN_{0s}}{d\Omega } &=&\frac{e^{2}\vartheta ^{2}}{4\pi \hbar c^{3}}%
\left[
\begin{array}{c}
\sin ^{2}\varphi  \\
\cos ^{2}\varphi
\end{array}
\right] \frac{\beta _{1}r_{s}^{^{\prime }2}\left| (\gamma ^{-2}-\chi
_{0}^{^{\prime }}+\vartheta ^{2})^{2}-\left| \beta _{1}\right|
r_{s}^{^{\prime }}\right| }{(\gamma ^{-2}-\chi _{0}^{^{\prime }}+\vartheta
^{2})^{2}}\frac{\omega _{B}L_{0}}{\sin ^{2}\theta _{B}}\cdot   \nonumber \\
&&\left| \frac{\exp (-i\omega _{B}L_{0}\frac{(\gamma ^{-2}-\chi
_{0}^{^{\prime }}+\vartheta ^{2})^{2}-\left| \beta _{1}\right|
r_{s}^{^{\prime }}}{2c(\gamma ^{-2}-\chi _{0}^{^{\prime }}+\vartheta
^{2})^{2}})}{\left| \beta _{1}\right| r_{s}^{^{\prime }}\exp (-i\omega
_{B}L_{0}\frac{(\gamma ^{-2}-\chi _{0}^{^{\prime }}+\vartheta
^{2})^{2}-\left| \beta _{1}\right| r_{s}^{^{\prime }}}{2c(\gamma ^{-2}-\chi
_{0}^{^{\prime }}+\vartheta ^{2})^{2}})-(\gamma ^{-2}-\chi _{0}^{^{\prime
}}+\vartheta ^{2})^{2}}\right| ^{2}.
\end{eqnarray}

We have investigated two geometries for PXR observation in Bragg case in
forward direction.

1. As one can see from (14),
maximum in angular distribution expression is reached
at condition
\begin{equation}
\omega _{B}L_{0}\frac{(\gamma ^{-2}-\chi _{0}^{^{\prime }}+\vartheta
^{2})^{2}-\left| \beta _{1}\right| r_{s}^{^{\prime }}}{2c(\gamma ^{-2}-\chi
_{0}^{^{\prime }}+\vartheta ^{2})^{2}}=2\pi n,\ \ n=0,1,...
\end{equation}
and the beats over polar angle \ $\vartheta $\ may be
observed. Obviously it is possible to choose a geometry where
these effects would be the most essential. Fig.2 demonstrates the
experimental scheme realizing the above condition.

\begin{figure}
\epsfxsize=10 cm
\centerline{\epsfbox{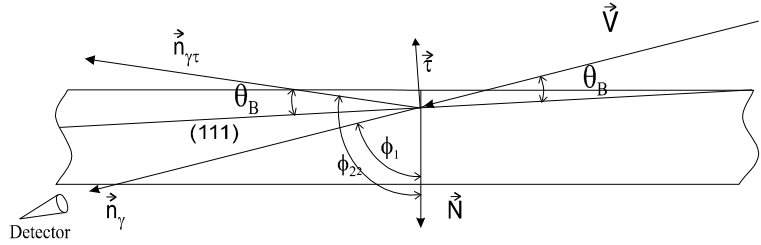}}
\caption{\it Geometry for observation of beating in PXR angular
distribution; $\gamma _{0}=\cos \phi _{1}, \gamma _{1}=\cos
\phi _{2}$.}
\end{figure}

We have carried out calculations for Si and LiH crystals.Experimental 
parameters
are given in table 1. All calculations
were performed for azimuth angle $\varphi =\pi /2$ (this corresponds to $%
\sigma $-polarization of PXR) and diffraction plane (111).

\begin{table} [t]
\vspace{0.2in}
\caption{Experimental parameters.}

\begin{center}
\begin{tabular} {|ccccccccc|c} \cline{1-9}
\footnotesize \it N & \footnotesize Crystal & \footnotesize Diffraction plane &
\footnotesize $\omega _{B}$,keV & \footnotesize$\gamma _{0}$ &
\footnotesize $\beta _{1}$ & \footnotesize $L_{0}$, cm & \footnotesize $L_{br}$, cm &
\footnotesize $L_{abs}$, cm &  \\ \cline{1-9}
\footnotesize 1 & \footnotesize Si & \footnotesize (111) & \footnotesize 30 &
\footnotesize 0,1211 & \footnotesize -11,57 & \footnotesize 8,26$\cdot $10$^{-2}$ &
\footnotesize 9,33$\cdot $10$^{-3}$ & \footnotesize 0,362 &  \\
\footnotesize 2 & \footnotesize Si & \footnotesize (111) & \footnotesize 30 &
\footnotesize 0,1177 & \footnotesize -8,43 & \footnotesize 8,50$\cdot $10$^{-2}$ &
\footnotesize 9,33$\cdot$10$^{-3}$ & \footnotesize 0,362 &  \\
\footnotesize 3 & \footnotesize Si & \footnotesize (111) & \footnotesize 30 &
\footnotesize 0,1142 & \footnotesize -6,55 & \footnotesize 8,76$\cdot $10$^{-2}$ &
\footnotesize 9,33$\cdot$10$^{-3}$ & \footnotesize  0,362 &  \\
\footnotesize 4 & \footnotesize Si & \footnotesize (111) & \footnotesize 30 &
\footnotesize 0,1055 & \footnotesize -4,03 & \footnotesize 9,48$\cdot $10$^{-2}$ &
\footnotesize 9,33$\cdot$10$^{-3}$ & \footnotesize 0,362 &  \\
\footnotesize 5 & \footnotesize Si & \footnotesize (111) & \footnotesize 30 &
\footnotesize 0,0882 & \footnotesize -2,02 & \footnotesize 1,13$\cdot $10$^{-1}$ &
\footnotesize 9,33$\cdot$10$^{-3}$ & \footnotesize 0,362 &  \\
\footnotesize 6 & \footnotesize Si & \footnotesize (111) & \footnotesize 30 &
\footnotesize 0,0659 & \footnotesize -1,00 & \footnotesize 1,52$\cdot $10$^{-1}$ &
\footnotesize 9,33$\cdot$10$^{-3}$ & \footnotesize 0,362 &  \\
\footnotesize 7 & \footnotesize Si & \footnotesize (111) & \footnotesize 30 &
\footnotesize 0,0349 & \footnotesize -0,36 & \footnotesize 2,87$\cdot $10$^{-1}$ &
\footnotesize 9,33$\cdot$10$^{-3}$ & \footnotesize 0,362 &  \\
\footnotesize 8 & \footnotesize Si & \footnotesize (111) & \footnotesize 20 &
\footnotesize 0,1636 & \footnotesize -4,85 & \footnotesize 3,06$\cdot $10$^{-3}$ &
\footnotesize 1,14$\cdot$10$^{-2}$ & \footnotesize 0,106 &  \\
\footnotesize 9 & \footnotesize Si & \footnotesize (111) & \footnotesize 20 &
\footnotesize 0,1636 & \footnotesize -4,85 & \footnotesize 3,06$\cdot $10$^{-2}$ &
\footnotesize 1,14$\cdot$10$^{-2}$ & \footnotesize 0,106 & \footnotesize * \\
\footnotesize 10 & \footnotesize LiH & \footnotesize (111) & \footnotesize 15 &
\footnotesize 0,3292 & \footnotesize -20,10 & \footnotesize 3,04$\cdot $10$^{-2}$ &
\footnotesize 4,60$\cdot$10$^{-2}$ & \footnotesize 17,4 &  \\
\footnotesize 11 & \footnotesize LiH & \footnotesize (111) & \footnotesize 20 &
\footnotesize 0,2507 & \footnotesize -25,35 & \footnotesize 3,99$\cdot $10$^{-2}$ &
\footnotesize 3,98$\cdot$10$^{-2}$ & \footnotesize 41,1 & \footnotesize * \\
\footnotesize 12 & \footnotesize LiH & \footnotesize (111) & \footnotesize 30 &
\footnotesize 0,1707 & \footnotesize -46,71 & \footnotesize 5,86$\cdot $10$^{-2}$ &
\footnotesize 3,25$\cdot$10$^{-2}$ & \footnotesize 139 &  \\ \cline{1-9}
\end{tabular}
\end{center}

\vspace{0.2in}
\end{table}

Fig.3 demonstrates spectral-angular distribution of parametric x-radiation
for experimental parameters shown in the 8-th row of  table 1. 
In this geometry
PXR is generated within frequency range where the coefficient characterizing
the efficiency of X-ray reflection by the atomic planes becomes nearly equal
to unit. In fig.3 the right part of spectrum is a flat curve - this is the
range of total Bragg reflection.

\begin{figure}
\epsfxsize = 10 cm
\centerline{\epsfbox{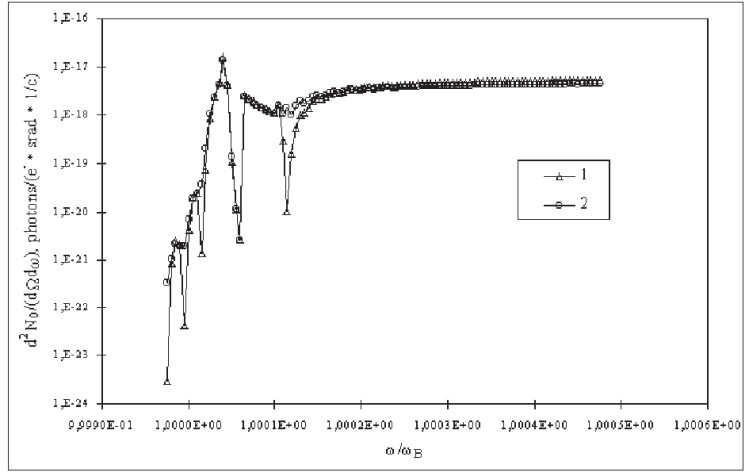}}
\caption{ \it Spectral-angular distribution of PXR for polar angle
$\vartheta =2,09$  mrad (azimuth angle $\varphi =\pi /2$).
1 - without multiple scattering (MS) taken into account,
2 - with MS.}
\end{figure}

The angular distribution for geometry described by the 3-d row of table 2
is shown in fig.4. The curve 1 is derived by
integration of spectral-angular distribution over the frequency range 
$\Delta \omega /\omega _{B}=10^{-4}$ within $\omega _{0}$,
corresponding to the center of PXR diffraction maximum for each pair $%
(\varphi ,\vartheta )$. The curve 2 is given by (14), it shows more
pronounced beating over polar angle than curve 1. 
The maxima of angular distribution for analytical
formula exceed the maximum values of distribution obtained by integration.
This can be explained by more exact account of absorption effect 
while performing
integration of spectral-angular distribution. Analytical formula was obtained
with some assumptions (neglecting of imaginary parts of crystal
susceptibilities) that leads to some
overestimation of PXR angular intensity.

\begin{figure}
\epsfxsize=10 cm
\centerline{\epsfbox{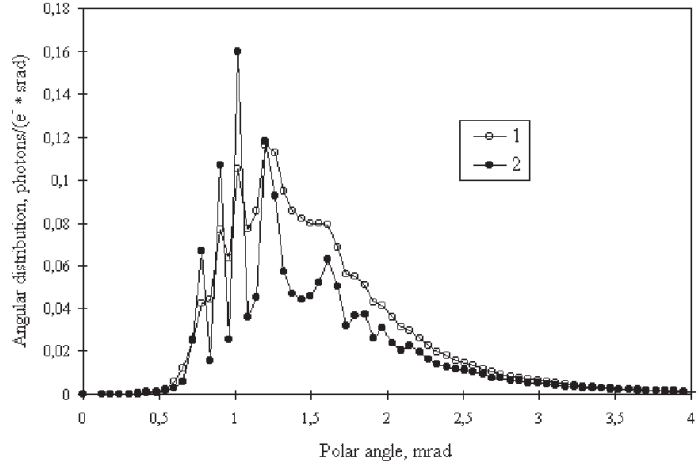}}
\caption{ \it Angular distribution of PXR derived on the base of:
1- integration of spectral-angular distribution over frequencies,
2 - analytical formula for angular distribution.}
\end{figure}

Fig.5 demonstrates angular distributions of PXR for different values of
geometrical factor $\beta _{1}$ -- first seven rows of table 1 (this
corresponds to different angles between (111) plane and surface of crystal
plate).

\begin{figure}
\epsfxsize=10 cm
\centerline{\epsfbox{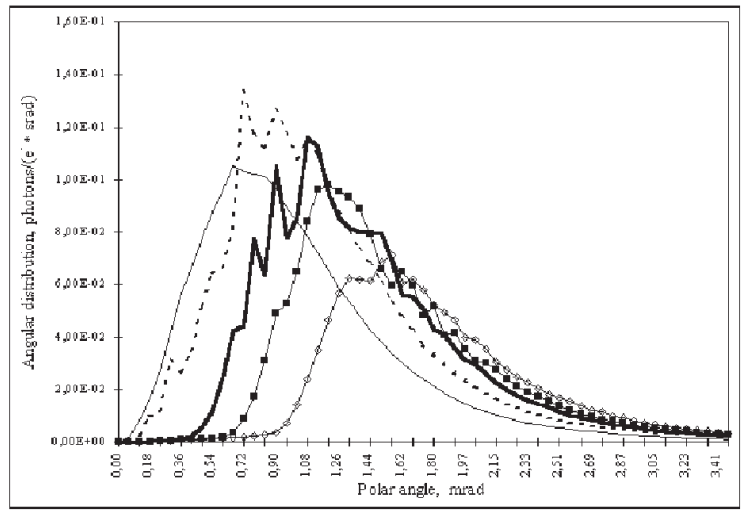}}
\caption {\it  PXR angular distributions for different values of $\beta
_{1}$ factor (from the left $\beta
_{1}=-2,02;-4,03;-6,55;8,43;-11,57$.}
\end{figure}

2. Another variant of experimental geometry is given by scheme shown in fig.
6. The Bragg angle $\theta _{B}$\ is equal to 90 degrees, this is 
the geometry
realizing normal incidence of particles on diffraction plane, the surface of
crystal plate is parallel to diffraction planes.

\begin{figure}
\epsfxsize=10 cm
\centerline{\epsfbox{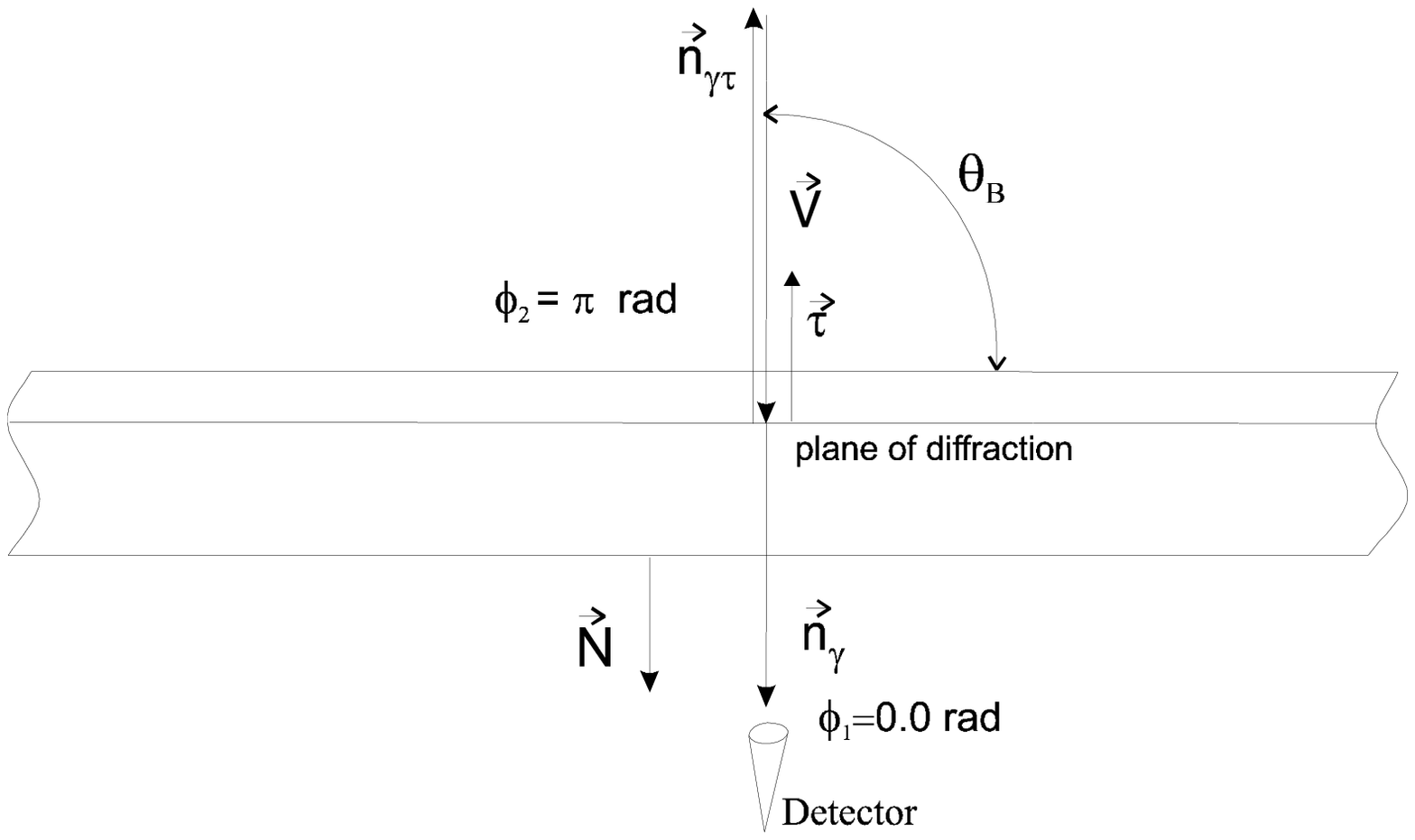}}
\caption{\it Experimental geometry realizing perpendicular incidence. }
\end{figure}

The calculations for the experimental parameters 
presented in table 2 were carried out .

\begin{table} [t]
\vspace{0.2in}
\caption{Experimental parameters.}
\begin{center}
\begin{tabular} {|ccccccccc|c} \cline{1-9}
\footnotesize \it N & \footnotesize Crystal & \footnotesize Diffraction plane &
\footnotesize $\omega _{B}$,keV & \footnotesize$\gamma _{0}$ &
\footnotesize $\beta _{1}$ & \footnotesize $L_{0}$, cm & \footnotesize $L_{br}$, cm &
\footnotesize $L_{abs}$, cm &  \\ \cline{1-9}
\footnotesize 1 & \footnotesize Si & \footnotesize (555) & \footnotesize9,887 &
\footnotesize 1,00 & \footnotesize -1,00 & \footnotesize 0,05 &
\footnotesize $1,63\cdot 10^{-2}$ & \footnotesize $1,28\cdot 10^{-2}$ &  \\
\footnotesize 2 & \footnotesize Si & \footnotesize (555) & \footnotesize9,887 &
\footnotesize 1,00 & \footnotesize -1,00 & \footnotesize 0,01 &
\footnotesize$1,63\cdot 10^{-2}$ & \footnotesize $1,28\cdot 10^{-2}$ & \footnotesize * \\
\footnotesize 3 & \footnotesize Si & \footnotesize (555) & \footnotesize 9,887 &
\footnotesize 1,00 & \footnotesize -1,00 & \footnotesize 0,005 &
\footnotesize $1,63\cdot 10^{-2}$ & \footnotesize $1,28\cdot 10^{-2}$ &  \\
\footnotesize 4 & \footnotesize LiH & \footnotesize (333) & \footnotesize 7,876 &
\footnotesize 1,00 & \footnotesize -1,00 & \footnotesize 0,05 &
\footnotesize $6,35\cdot 10^{-2}$ & \footnotesize 2,52 & \footnotesize* \\
\footnotesize 5 & \footnotesize LiH & \footnotesize (333) & \footnotesize 7,876 &
\footnotesize 1,00 & \footnotesize -1,00 & \footnotesize 0,01 &
\footnotesize $6,35\cdot 10^{-2}$ & \footnotesize 2,52 & \\
\footnotesize 6 & \footnotesize LiH & \footnotesize (333) & \footnotesize 7,876 &
\footnotesize 1,00 & \footnotesize -1,00 & \footnotesize 0,005 &
\footnotesize $6,35\cdot 10^{-2}$ & \footnotesize 2,52 & \\ \cline{1-9}
\end{tabular}
\end{center}
\vspace{0.3in}
\end{table}
The spectral-angular distribution for geometry corresponding to the 3-d
row of table 2 is shown in fig.7.

\begin{figure}
\epsfxsize = 10 cm
\centerline{\epsfbox{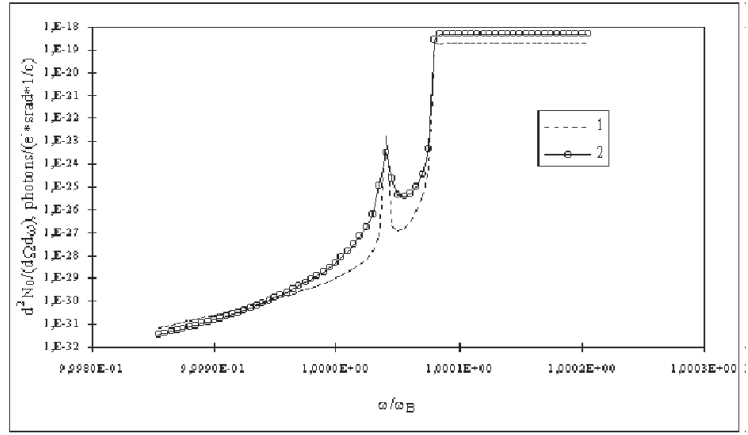}}
\caption{\it Spectral-angular distributions of PXR for polar angle
$\vartheta=12,23 mrad$ (azimuth angle) $\varphi=\pi /2 $
1 -- without account of MS; 2 -- with account of MS.}
\end{figure}

\begin{figure}
\epsfxsize=10 cm
\centerline{\epsfbox{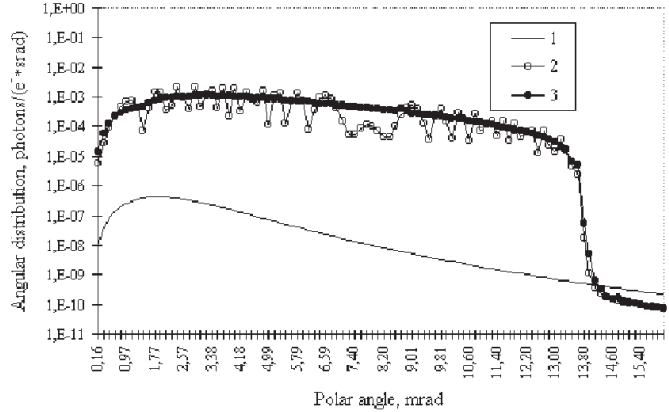}}
\caption{\it Angular distributions of PXR obtained on the base of: 1 -
analytical formula for angular intensity, 2 - integrating spectral-angular
distribution over frequencies ($\Delta \omega /\omega _{B}=10^{-4}$)
without taking into account multiple scattering, 3 - integrating
spectral-angular distribution over frequencies ($\Delta \omega /\omega
_{B}=10^{-4}$) taking into account multiple scattering.  }
\end{figure}

Fig.8 presents PXR angular distribution corresponding to the 3-d row of
table 2. Such essential difference between results obtained on the base of
analytical formula and that of integration of spectral-angular
distribution can be explained by the fact that the formula (14) does not take
into account the nonhomogeneous waves arising at the total reflection in
Bragg case (in contrast with Laue geometry where the total reflection 
does not realized) (see fig. 7 -- the flat part
of spectral-angular distribution curve).

\section{PXR at large angles with respect to particle's velocity.}

As well as for forward PXR, in order to obtain angular
distribution for diffracted PXR (see fig.1, (b)) taking into account
the effect of absorption, possible total reflection of quanta in a crystal
and contribution of both slow and fast
waves in radiation intensity it is necessary to perform the numerical
integration of spectral-angular distribution (12) over frequencies in the
vicinity of $\omega _{B}$. We have carried out calculations for the geometry
presented in fig.9.

\begin{figure}
\epsfxsize=10 cm
\centerline{\epsfbox{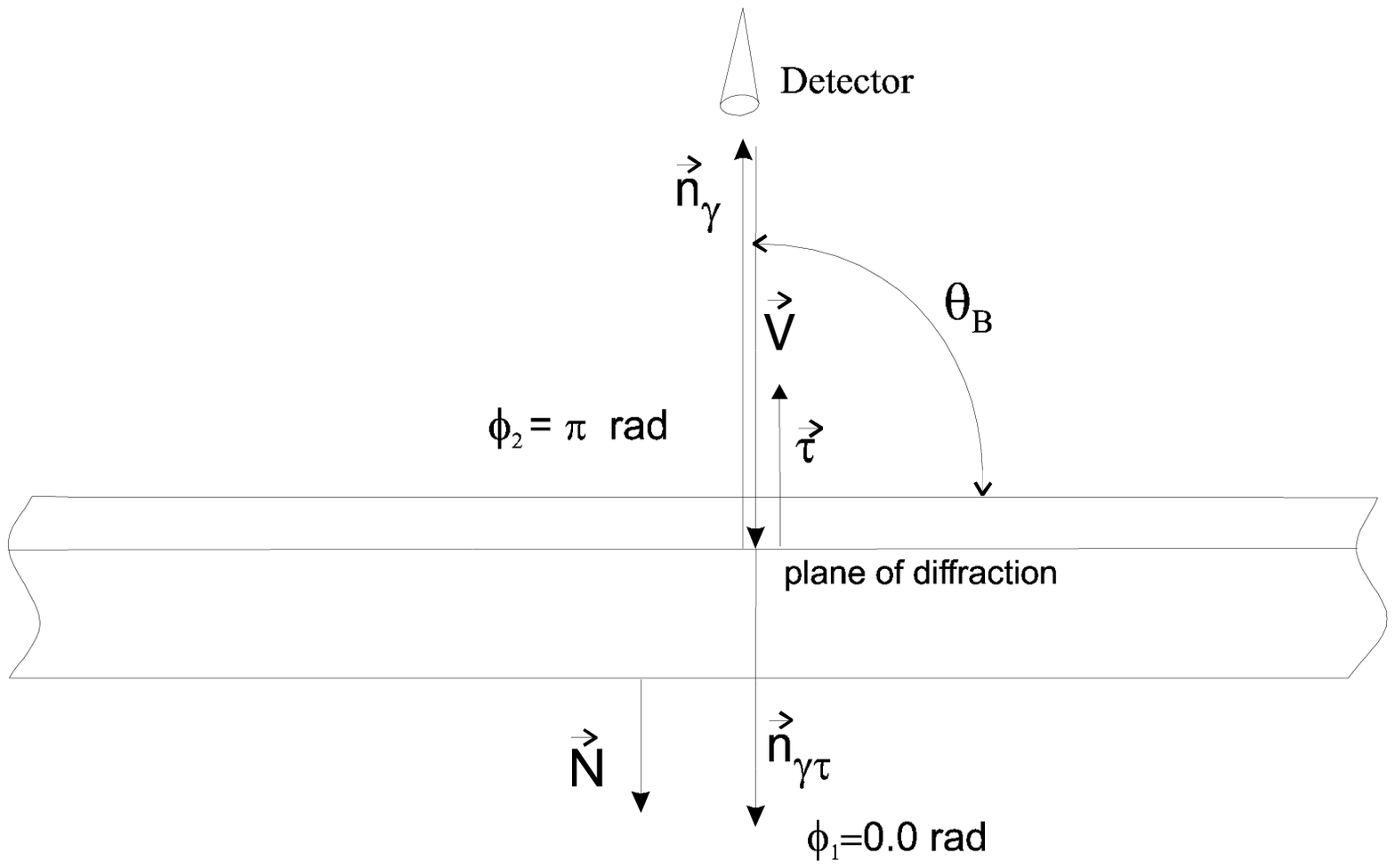}}
\caption{\it Geometry of backward diffraction. }
\end{figure}

\begin{figure}
\epsfxsize=10 cm
\centerline{\epsfbox{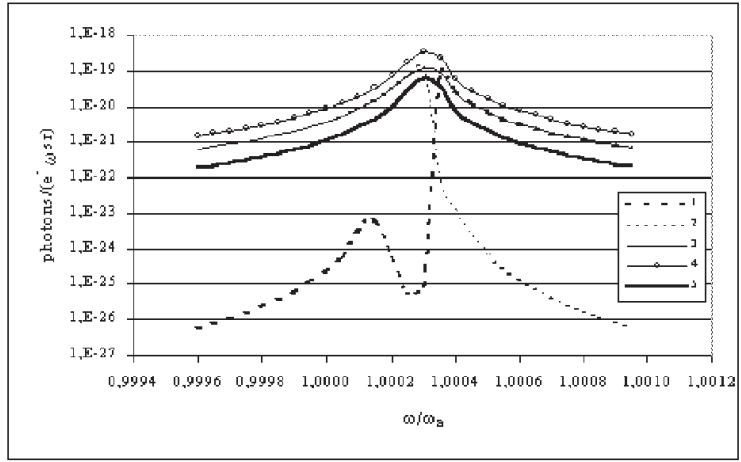}}
\caption {\it Spectral-Angular distribution for backward diffraction
in Si crystal, diffraction plane (111), particle's energy 86,9 MeV,
$\vartheta =18,6$ mrad (azimuth angle $\varphi =\pi /2$ ),
$\omega_{B}=1,98 keV$, $L=1,25\cdot 10^{-3}$ cm. 1 -- PXR spectral-angular
distribution (SAD) for the first dispersion
branch $\mu =1$ without the terms proportional to $\frac{1}
{\omega -\vec{k}\vec{v}}$;
2 -- SAD for $\mu =2$ without the terms proportional to $\frac{1}{\omega -\vec{k}\vec{v}}$;
3 -- SAD for the both dispersion branches without the terms $\frac{1}{\omega -\vec{k}\vec{v}}$;
4 -- SAD for the both dispersion branches without the terms proportional
to $\frac{1}{\omega -\vec{k}_{\mu s}\vec{v}}$;
5 -- SAD described by the formula (12).}
\end{figure}

\begin{figure}
\epsfxsize=10 cm
\centerline{\epsfbox{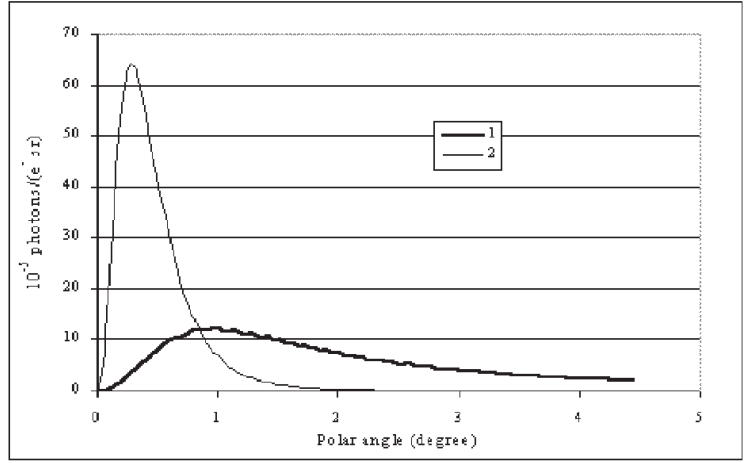}}
\caption {\it Angular distribution for backward diffraction, derived by
integration (12) over frequencies in the range $\Delta \omega /\omega
_{B}=10^{-3}$ in silicon crystal for azimuth angle $\varphi =\pi
 /2 $, diffraction plane (111), particle's energy 86,9 MeV, $\omega
_{B}=1,98$ keV, $L=1,25\cdot 10^{-3}$ cm.
1-- PXR angular distribution for the both dispersion branches
without integration the terms proportional to $\frac{1}{\omega -\vec{k}\vec{v}}$;
2-- PXR angular distribution with taking into account during
integration all terms in expression (12).}
\end{figure}

\begin{figure}
\epsfxsize=10 cm
\centerline{\epsfbox{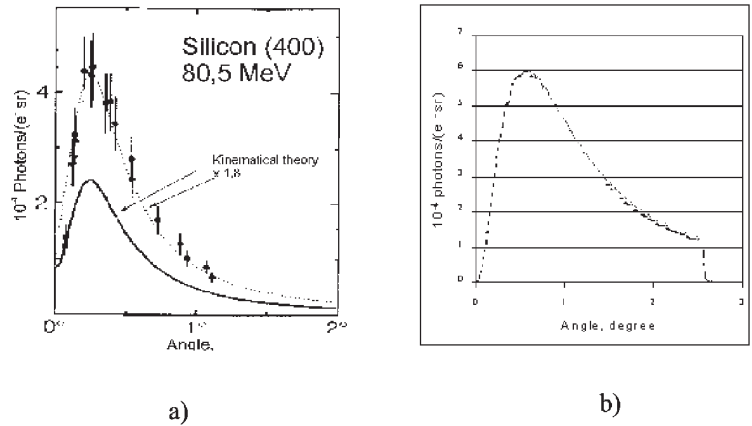}}
\caption{\it Angular distribution for backward diffraction, in silicon
crystal for azimuth angle $\varphi =\pi /2$, diffraction plane
(400), particle's energy 80,5 MeV, $\omega _{B}=4,57$ keV,
$L=1,25\cdot 10^{-3}$ cm. a -- results of \cite{19};
b -- derived by integration (12) over frequencies in the range
$\Delta \omega /\omega _{B}=10^{-3}$}
\end{figure}

Figs. 10 and 11 show spectral-angular and angular distributions for backward 
diffraction. 
How one can see from Fig.12, expression (12), taking into account effects
of a dynamic diffraction, allows to describe  the experiment more precisely
than the formula of the kinematic theory \cite{15}.

\section{Effect of multiple scattering on PXR.}

In addition to absorption of photons in crystalline plate
multiple scattering of charged particles by
atoms of crystal is the other
essential factor also leading to limitation of longitudinal dimensions of
PXR generation range. So, the PXR characteristics strongly depend on
relation between crystal thickness $L_{0}$ and coherent length of
bremsstrahlung radiation $L_{br}$. When $L_{0}<<L_{br}$ the
influence of multiple scattering reduces actually to a small addition to PXR
intensity due to bremsstrahlung radiation mechanism. In the opposite case,
when $L_{0}>L_{br}$, multiple scattering significantly changes PXR parameters. 
The calculations taking into account multiple scattering effect
were carried out on the base of expressions given in \cite{7}.

The relation  between $L_{br}$ and crystal thickness $L_{0}$ determines
the influence  of multiple scattering on  PXR characteristics
(the values $L_{br}$ are contained in the 8-th columns of tables 1 and 2)
Figs. 3, 7, 8 represents
PXR (spectral-angular and angular) distributions taking into account
multiple scattering influence. 
All these distributions were obtained for thin crystalline plates
($L$ $<L_{br}$).

In fig.13 angular distributions of PXR and bremsstrahlung radiation  
caused by multiple scattering for the parameters presented in the 3-d row
of table 1 are shown. 
These distributions were obtained by integration of
spectral-angular distributions while taking multiple scattering into account
and without it. This is the case of intense MS, $L_{0}\sim 9$ $L_{br}$, MS
leads to strong reduction of PXR intensity.

\bigskip

\begin{figure}
\epsfxsize = 10 cm
\centerline{\epsfbox{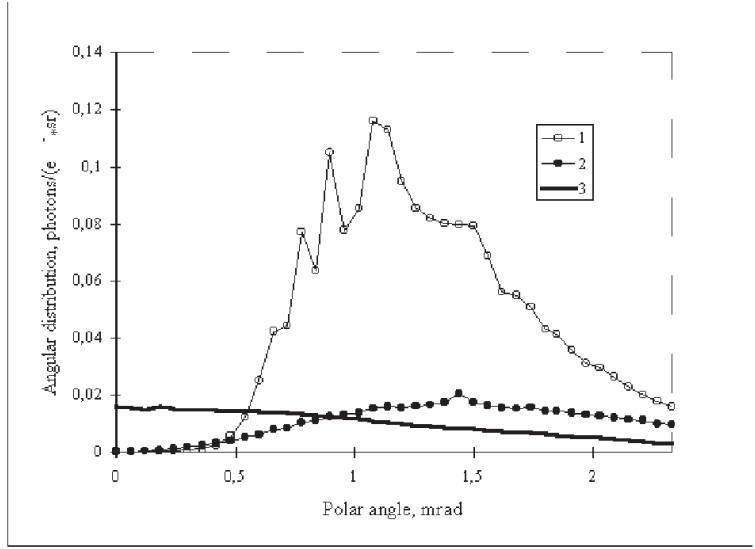}}
\caption {\it PXR angular distribution in silicon crystal:
1 -- PXR without account of MS; 2 -- PXR with account of MS; 3 --
Bremsstrahlung caused by MS.}
\end{figure}

\begin{figure}
\epsfxsize=10 cm
\centerline{\epsfbox{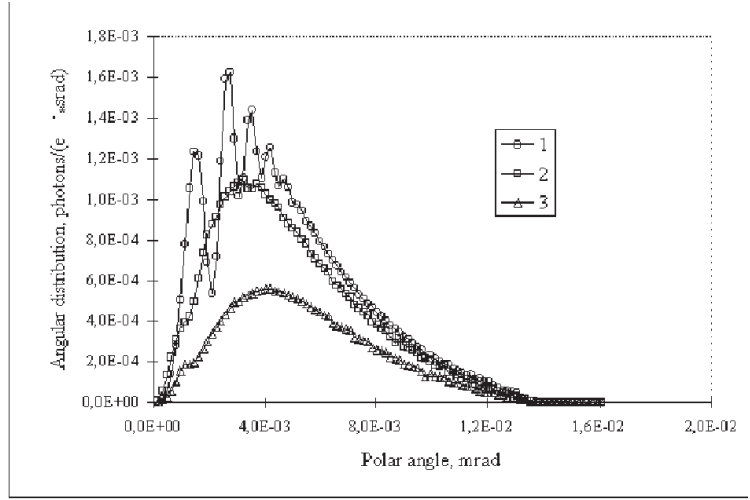}}
\caption{\it PXR angular distribution in silicon crystal with account of
MS: 1 -- $L=0,005$ cm, 2 -- $L=0,01$ cm, 3 -- $L=0,05 $ cm. }
\end{figure}

Fig.14 demonstrates dependence of intensity of angular distribution on the
thickness of crystalline plate. The parameters describing these situations
are given in rows 1-3 of table 2. For plate of thickness $L_{0}=0,005$ cm MS
is not very essential (see fig. 7), for thickness of 0,01 cm ($L_{0}\sim
L_{br}$) MS is already noticeable, and in plate with the thickness of 0,05
cm MS plays the important role in PXR intensity reduction.

\section{PXR in LiH crystall.}

Use in experiments of LiH crystal enables to realize weak MS
even for sufficiently thick crystals (see rows 11 and 12 of table 1 and rows
4, 5, 6 of table 2). In case of Bragg diffraction for geometry described by
fig.2 it is possible to observe beating over polar angle in angular
distributions measurements. The experimental geometries more advantageous
for PXR observation are marked in table 1 by (*). For
these experimental schemes the comparison between angular intensities of 
PXR and
TXR was carried out ($\Delta \omega /\omega _{B}=10^{-4}$). Figure 15
demonstrates angular distributions of PXR and TXR for LiH crystal.

For Bragg diffraction in case of normal incidence the geometries preferable
for PXR observation possibility are asterisked at margins
of table 2.

Fig.16 demonstrates comparison of PXR and TXR angular intensities for
situation described by the 4-th row of table 2.

\begin{figure}
\epsfxsize=10 cm
\centerline{\epsfbox{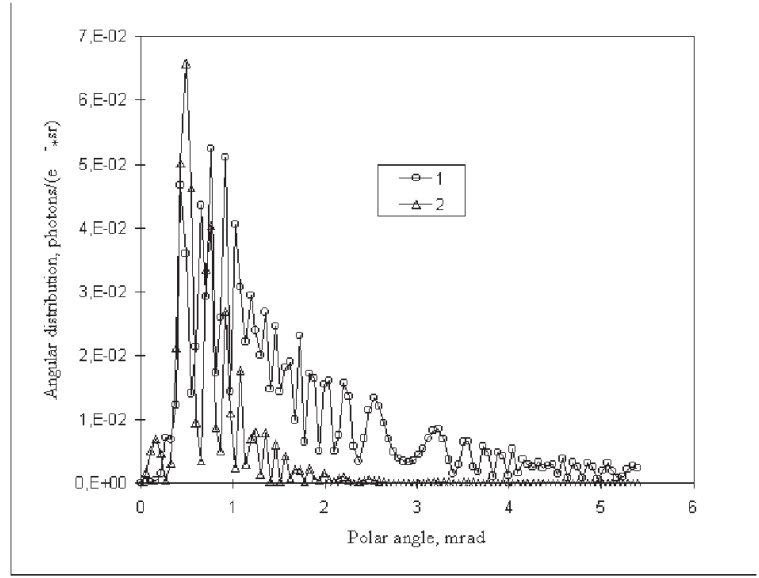}}
\caption{\it Angular distribution in LiH crystal for geometry described
by fig.2:
1 --  PXR, 2 -- TXR.}
\end{figure}

\begin{figure}
\epsfxsize=10 cm
\centerline{\epsfbox{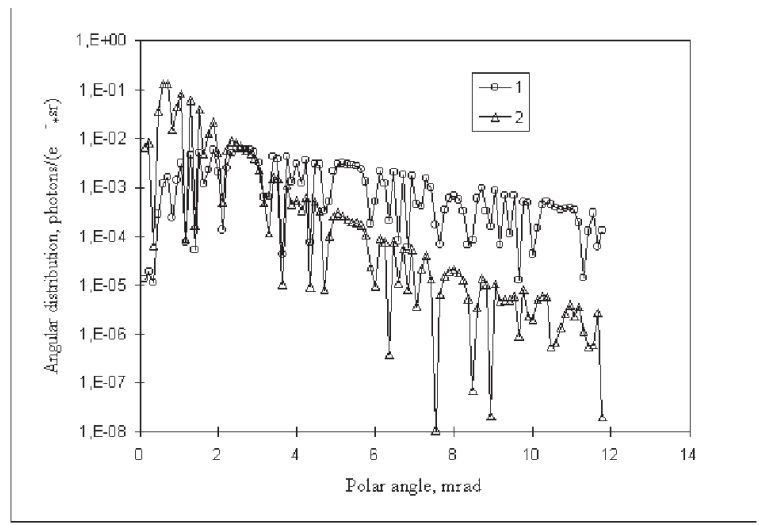}}
\caption{\it Angular distribution in LiH crystal, case of normal
incidence geometry: 1 - PXR, 2 - TXR. $\omega _{B}=7,88$ keV,
$L_{0}=0,01$ cm.}
\end{figure}

\section*{Conclusion.}

The only precision account of all dynamic effects (total
Bragg reflection, interference of slow and fast waves or, by the other words, taking into
consideration of all terms in (12), allows to obtain the real picture
of diffraction radiation in case of Bragg geometry.
The contributions of both the slow and the fast 
waves in radiation  intensity for Bragg geometry  are comparable and only their joint 
account allows to describe the experimental results.

\end{document}